\documentclass[12pt]{article}
\usepackage{a4wide}
\usepackage{amssymb}
\usepackage[caption=false]{subfig}
\usepackage{graphicx}
\usepackage{amsmath}
\usepackage{amsfonts}
\usepackage{mathtools}
\usepackage{color}
\usepackage{bm}
\usepackage{mathrsfs}
\usepackage{epstopdf}
\usepackage{url}
\usepackage{footnote}
\usepackage{textcomp}
\usepackage{dsfont}
\usepackage{ulem}
\usepackage{hyperref}
\usepackage{cite}
\usepackage{enumerate}

\begin{document}
{\renewcommand{\thefootnote}{\fnsymbol{footnote}}
\begin{center}
{\LARGE  Testing loop quantum gravity from observational consequences of non-singular rotating black holes}\\
\vspace{1.5em}
Suddhasattwa Brahma$^1$\footnote{e-mail address: {\tt
		suddhasattwa.brahma@gmail.com}}, Che-Yu Chen$^2$\footnote{e-mail address: {\tt
		b97202056@gmail.com}}
 and Dong-han Yeom$^{3,4}$\footnote{e-mail address: {\tt
 		innocent.yeom@gmail.com}}
\\
\vspace{0.5em}
$^1$ Department of Physics, McGill University, Montr\'{e}al, QC H3A 2T8, Canada\\
\vspace{0.5em}
$^2$ Institute of Physics, Academia Sinica, Taipei 11529, Taiwan\\
\vspace{0.5em}
$^3$ Department of Physics Education, Pusan National University, Busan 46241, Korea\\
\vspace{0.5em}
$^4$ Research Center for Dielectric and Advanced Matter Physics,\\ Pusan National University, Busan 46241, Korea
\vspace{1.5em}
\end{center}
}

\setcounter{footnote}{0}

\begin{abstract}
  The lack of rotating black hole models, which are typically found in nature, in loop quantum gravity (LQG) substantially hinders the progress of testing LQG from observations. Starting with a non-rotating LQG black hole as a seed metric, we construct a rotating spacetime using the revised Newman-Janis algorithm. The rotating solution is non-singular everywhere and it reduces to the Kerr black hole asymptotically. In different regions of the parameter space, the solution describes \textit{i)} a wormhole without event horizon (which, we show, is almost ruled out by observations), \textit{ii)} a black hole with a spacelike transition surface inside the event horizon, or \textit{iii)} a black hole with a timelike transition region inside the inner horizon. It is shown how fundamental parameters of LQG can be constrained by the observational implications of the shadow cast by this object. The causal structure of our solution depends crucially only on the spacelike transition surface of the non-rotating seed metric, while being agnostic about specific details of the latter, and therefore captures universal features of an effective rotating, non-singular black hole in LQG.
\end{abstract}

\section{Introduction}
Direct detection of gravitational waves and images of black hole shadows ushers in a golden era of black hole astronomy. At present, these extreme stellar objects serve as our best candidates for testing fundamental quantum gravity theories, such as loop quantum gravity (LQG). LQG, being a non-perturbative approach to quantum gravity, goes beyond general relativity to resolve classical singularities in cosmological and (non-rotating) black hole spacetimes and, in this work, we extend similar techniques to the case of the rotating Kerr-like black hole. Indeed, a consistent LQG black hole (LQGBH) model should not only provide a singularity-free description of the spacetime inside the horizon, it must also have a viable picture for the exterior region with verifiable consequences for these observations.

Due to technical difficulties in solving the LQG equations of motion, especially when using real-valued Ashtekar-Barbero variables, axisymmetric spacetimes have remained largely unexplored. Since this is the class of spacetimes to which the Kerr black hole belongs, therefore, a direct loop quantization of rotating black holes is yet to be achieved (see \cite{LQG_Axi1, LQG_Axi2} for previous attempts). However, from the point of view of phenomenology, this is the primary case of interest since most of the astrophysical black holes which have been observed are those with non-zero angular momenta.

On the other hand, LQG effective equations have been thoroughly investigated for static, spherically symmetric, and non-rotating spacetimes, resulting in quantum extensions of the Schwarzschild black hole (see \cite{Bodendorfer:2019nvy, Bodendorfer:2019jay, LQG1, LQG2, LQG3, LQG4, LQG5, LQG6, LQG7, LQG8, LQG9, LQG10, LQG11, LQG12, Ashtekar:2018lag,Barrau:2018rts} for an incomplete list of these models, \cite{LQGBH_review} for a critical review and \cite{SigChange1,SigChange2} for signature-changing solutions). In this letter, starting from a non-rotating LQGBH \cite{Bodendorfer:2019nvy,Bodendorfer:2019jay}, we construct a rotating spacetime using the Newman-Janis-Algorithm (NJA) \cite{Newman:1965tw}. As a solution-generating method, NJA is successful in constructing the Kerr (Kerr-Newman) solution from the Schwarzschild (Reissner-Nordstr\"om) black hole. We wish to follow a similar strategy for their (loop) quantum counterparts in the hope that such a solution will not only exhibit a non-singular geometry that one  expects, but also tell us how LQG effects can be tested in a realistic manner. A priori, it might seem a little \textit{ad hoc} to construct LQG solutions of rotating black holes in this way. However, this is similar in spirit to the ``effective equations'' one typically employs in symmetry-reduced models of LQG (\textit{e.g.} for LQGBHs), which include non-perturbative corrections inspired from the full theory. Analogously, we derive an \textit{effective} rotating, singularity-free spacetime which captures key aspects of LQG.

Previous attempts at generating rotating spacetimes, using NJA, starting from a non-rotating LQGBH, suffer either from using the now-outdated self-dual variables formalism \cite{Liu:2020ola} or an incorrect implementation of NJA \cite{Caravelli:2010ff,AzregAinou:2011fq}. The non-rotating LQGBH \cite{Bodendorfer:2019nvy, Bodendorfer:2019jay} that we are going to consider as the seed metric has several attractive features: In addition to the resolution of classical singularities as is expected to happen in LQG, the quantum effects (quantified by a single parameter) rapidly die out when moving away from the center, with a well-defined asymptotic region in the exterior, a property not shared by all effective models of LQGBHs \cite{Bouhmadi-Lopez:2019hpp}. We will show that the rotating counterpart also retains these characteristics. However, note that our solution is more general and some of its crucial features do not depend on explicit details of the seed metric we have used, thereby capturing some universal properties of rotating LQGBHs.

As we will show, the inclusion of spin naturally enriches the spacetime structure. In particular, it is possible to generate a rotating wormhole without horizon, although this geometry is disfavored by the measurement of the shadow of M87* by the Event Horizon Telescope Collaboration (EHT) \cite{Akiyama:2019cqa}. The most intriguing geometry is a regular black hole containing two horizons, with a timelike transition surface inside the inner one. Such a geometry is observationally favored by the requirement that the quantum parameter is extremely small. 

\section{Non-rotating LQGBH} 
On solving the LQG effective equations, the quantum extension of the Schwarzschild metric reads  \cite{Bodendorfer:2019nvy,Bodendorfer:2019jay}
\begin{equation}
ds^2=-\tilde{a}(x)d\tau^2+\frac{dx^2}{\tilde{a}(x)}+b(x)^2d\Omega_2^2\,.\label{BMMmetric}
\end{equation}
The metric functions are defined in terms of the radial variable $x\in(-\infty,\infty)$ as
\begin{align}
b(x)^2=&\,\frac{A_\lambda}{\sqrt{1+x^2}}\frac{M_B^2\left(x+\sqrt{1+x^2}\right)^6+M_W^2}{\left(x+\sqrt{1+x^2}\right)^3}\,,\label{BMM2metrica1}\\
\tilde{a}(x)=&\,\left(1-\sqrt{\frac{1}{2A_\lambda}}\frac{1}{\sqrt{1+x^2}}\right)\frac{1+x^2}{b(x)^2}\,,\label{BMM2metricfunctions}
\end{align}
where $M_B$ and $M_W$ correspond to two Dirac observables in the model. For convenience, we have defined a dimensionless parameter $A_\lambda\equiv(\lambda_k/M_BM_W)^{2/3}/2$, where the quantum parameter $\lambda_k$ originates from holonomy modifications \cite{Bodendorfer:2019nvy,Bodendorfer:2019jay}. In LQG, these non-perturbative corrections arise from regularizing the curvature components when considering holonomies around loops which can only be shrunk to the minimum non-zero eigenvalue of the area-operator (known as the \textit{area gap}), as opposed to taking the limit to zero as in classical general relativity. One of our main findings is that the quantum parameter, and thereby this fundamental `area-gap', is constrained by observations of shadows of rotating black holes.

The most important feature of this LQGBH \eqref{BMMmetric} is that inside the black hole, the areal radius $b$ reaches a minimum value, representing a \textit{spacelike transition surface} which smoothly connects an asymptotically Schwarzschild black hole to a white hole with mass $M_B$ and $M_W$, respectively. Specifically, we will focus on the physically interesting case of the symmetric bounce in which $M_B=M_W$, \textit{i.e.} the spacetimes are symmetric with respect to the transition surface ($x=0$). Rescaling the coordinates $(x,\tau)\rightarrow(y,t)$ as $y\equiv\sqrt{8A_\lambda}M_Bx$ and $t\equiv\tau/\sqrt{8A_\lambda} M_B$, the metric \eqref{BMMmetric} can be rewritten as
\begin{equation}
ds^2=-8A_\lambda M_B^2\tilde{a}(y)dt^2+\frac{dy^2}{8A_\lambda M_B^2\tilde{a}(y)}+b(y)^2d\Omega_2^2\,.\label{BMMmetricformal1}
\end{equation}
When $|y|\rightarrow\infty$, we have $|y|\rightarrow b$ and $8A_\lambda M_B^2\tilde{a}(y)\rightarrow1-2M_B/b$. Therefore, the metric \eqref{BMMmetricformal} reduces to the Schwarzschild one in the asymptotic limit ($b\rightarrow\infty$).

\section{Rotating LQGBH}

The rotating counterpart of \eqref{BMMmetric} is obtained using NJA, in which the spin parameter $a$ is included through a complex shift on the advanced null coordinates \cite{Newman:1965tw}. In particular, we use the revised NJA \cite{Azreg-Ainou:2014pra} which allows a valid representation of the resulting metric in the Boyer-Lindquist coordinate system $(t,y,\theta,\varphi)$. As a result, the metric of the rotating LQG compact object (rLQGO){\footnote{We use \textit{compact object} instead of {\textit{black hole} because, as will be shown later, the resulting spacetime could be without any trapping regions, in some parts of the parameter space.}} can be cast in a Kerr-like form (see the Appendix for details of the construction)
\begin{align}
ds^2=&-\left(1-\frac{2Mb}{\rho^2}\right)dt^2-\frac{4aMb\sin^2\theta}{\rho^2}dtd\varphi\nonumber\\&+\rho^2d\theta^2+\frac{\rho^2dy^2}{\Delta}+\frac{\Sigma\sin^2\theta}{\rho^2}d\varphi^2\,,\label{finalrot}
\end{align}
where $\rho^2=b^2+a^2\cos^2\theta$, $M=b\left(1-8A_\lambda M_B^2\tilde{a}\right)/2$, $\Delta=8A_\lambda M_B^2\tilde{a}b^2+a^2$, and $\Sigma=\left(b^2+a^2\right)^2-a^2\Delta\sin^2\theta$. Note that the functions $\tilde{a}$, $b$, $M$, and $\Delta$ are functions of $y$, as can be seen from Eqs.~\eqref{BMM2metrica1} and \eqref{BMM2metricfunctions}. 

Firstly, note that the metric \eqref{finalrot} reduces to Kerr asymptotically for $|y|\rightarrow\infty$, recovering the expected classical limit, while in the limit $a\rightarrow0$, the static LQGBH \eqref{BMMmetricformal1} is regained. Furthermore, setting $(a, M)\rightarrow 0$ gives one the flat limit \cite{Gan:2020dkb}, satisfying an essential consistency check lacking in some quantum gravity inspired solutions \cite{Hossenfelder:2009fc}. Secondly, $\Delta=0$ defines the event horizon of rLQGO, where the variable $y_h$  satisfies (see Eqs.~\eqref{BMM2metrica1} and \eqref{BMM2metricfunctions}):
\begin{equation}
		\sqrt{8A_\lambda+\left(y_h^2/M_B^2\right)}=1\pm\sqrt{1-\left(a^2/M_B^2\right)}\,,\label{ehEX}
\end{equation}
with the plus (minus) sign indicating the outer (inner) horizon on each side of the transition surface. The expression under the radical on the r.h.s. of Eq.~\eqref{ehEX} implies that there is a maximum spin for rLQGO: $|a|\le M_B$, which is the same as the Kerr bound. Evidently, the spacetime structure of rLQGO strongly depends on the values of the parameters $\{a,A_\lambda\}$ under consideration. As illustrated in Figure~\ref{fig.boundary}, the transition surface can either be outside the outer horizon (region I), or between the two horizons (region II), or inside the inner horizon (region III). These regions are split by the boundaries which denote the case when the transition surface is on the outer (red curve) and the inner (blue curve) horizons.

\begin{figure}[t]
\center\includegraphics[scale=0.8]{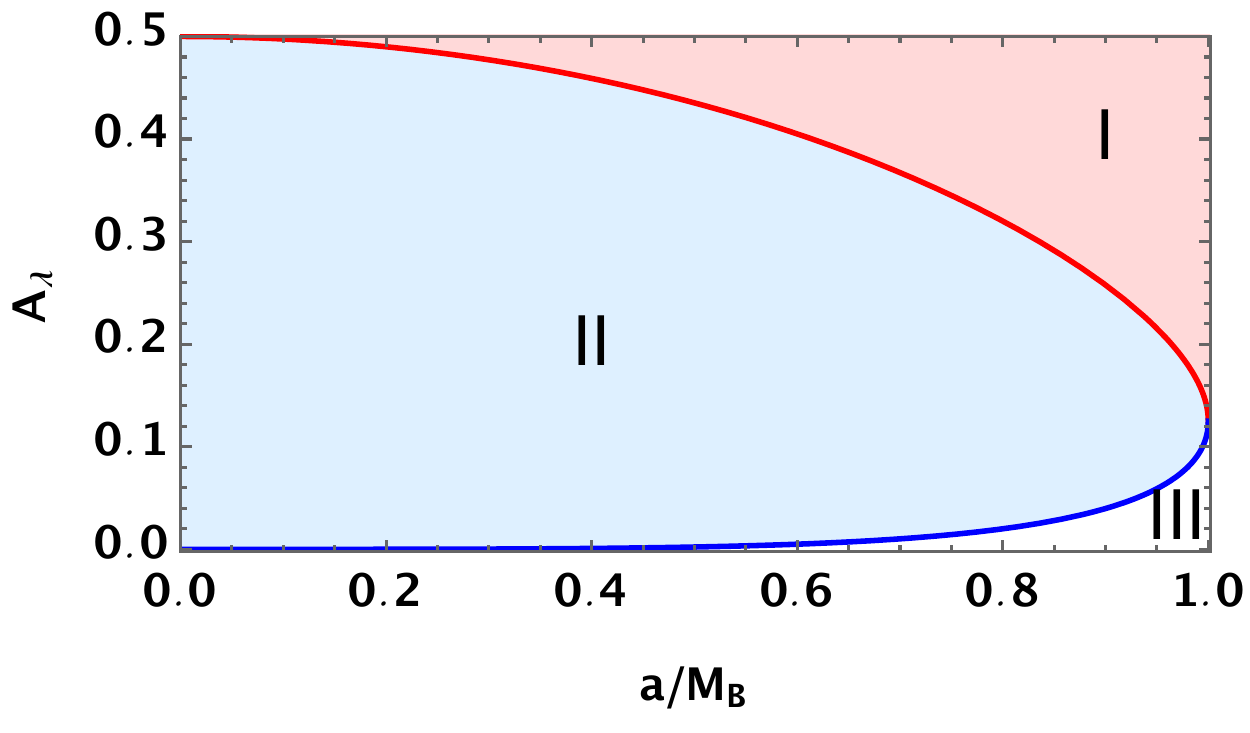}
\caption{\label{fig.boundary}The spacetime structure of the rLQGO metric \eqref{finalrot} with respect to the parameter space $\{a,A_\lambda\}$. In regions I, II, and III, the transition surface $(y=0)$ is located outside the outer horizon, between the inner and outer horizons, and inside the inner horizon, respectively. On the red (blue) curve, the transition surface is located on the outer (inner) horizon.}
\end{figure}

\textbf{\textit{Region I}} --  In this region of parameter space, the rLQGO is a rotating wormhole (Figures~\ref{fig.emb}(a) and \ref{fig.penrose1}(a)) without horizon. Its spacetime structure resembles that of the phenomenological Kerr-like wormhole proposed in Ref.~\cite{Bueno:2017hyj}. However, the Arnowitt-Deser-Misner mass of rLQGO \eqref{finalrot} is always $M_B$ (see the Appendix), while that of the model in Ref.~\cite{Bueno:2017hyj} depends explicitly on the throat parameter. Note that the ringdown signals generated by this type of wormholes are characterized by echos \cite{Bueno:2017hyj}.

\textbf{\textit{Region II}} -- The transition surface is hidden behind the outer event horizon and becomes spacelike (Figure~\ref{fig.emb}(b)). The green region is inside the event horizon where $t$ and $y$ exchange roles and the transition surface is located at the narrowest point in the middle. As expected, the Penrose diagram for this type of rotating black holes is similar to that of its non-rotating counterpart \cite{Bodendorfer:2019nvy,Bodendorfer:2019jay} (Figure~\ref{fig.penrose1}(b)), rendering the inner horizon irrelevant. This is because as $a/M_B \rightarrow 0$, the rLQGO tends to be in  region II, as long as the transition surface is hidden by the outer horizon.

\textbf{\textit{Region III}} -- Given a non-zero finite value of $a/M_B$, this region is characterized by a small $A_\lambda$ and, thus, is the most physically relevant one for considering rotating black holes. The classical ring singularity behind the Cauchy horizon of the Kerr black hole is replaced by a timelike transition surface. As shown in its Penrose diagram (Figure~\ref{fig.penrose3}), a timelike trajectory (the black dashed curve) entering the black hole crosses the inner horizons (blue solid lines). Thereafter, this trajectory can be extended into another universe either by going upwards (trajectory A) without touching the transition surface (blue dashed lines), or by crossing the transition surface into another interior patch (trajectory B). Its embedding diagram (Figure~\ref{fig.emb}(c)) has to terminate at a timelike surface outside the transition surface since it cannot be extended vertically downward any further (the surface of the cone becomes horizontal).

\begin{figure}[t]
\center(a)\includegraphics[scale=0.35]{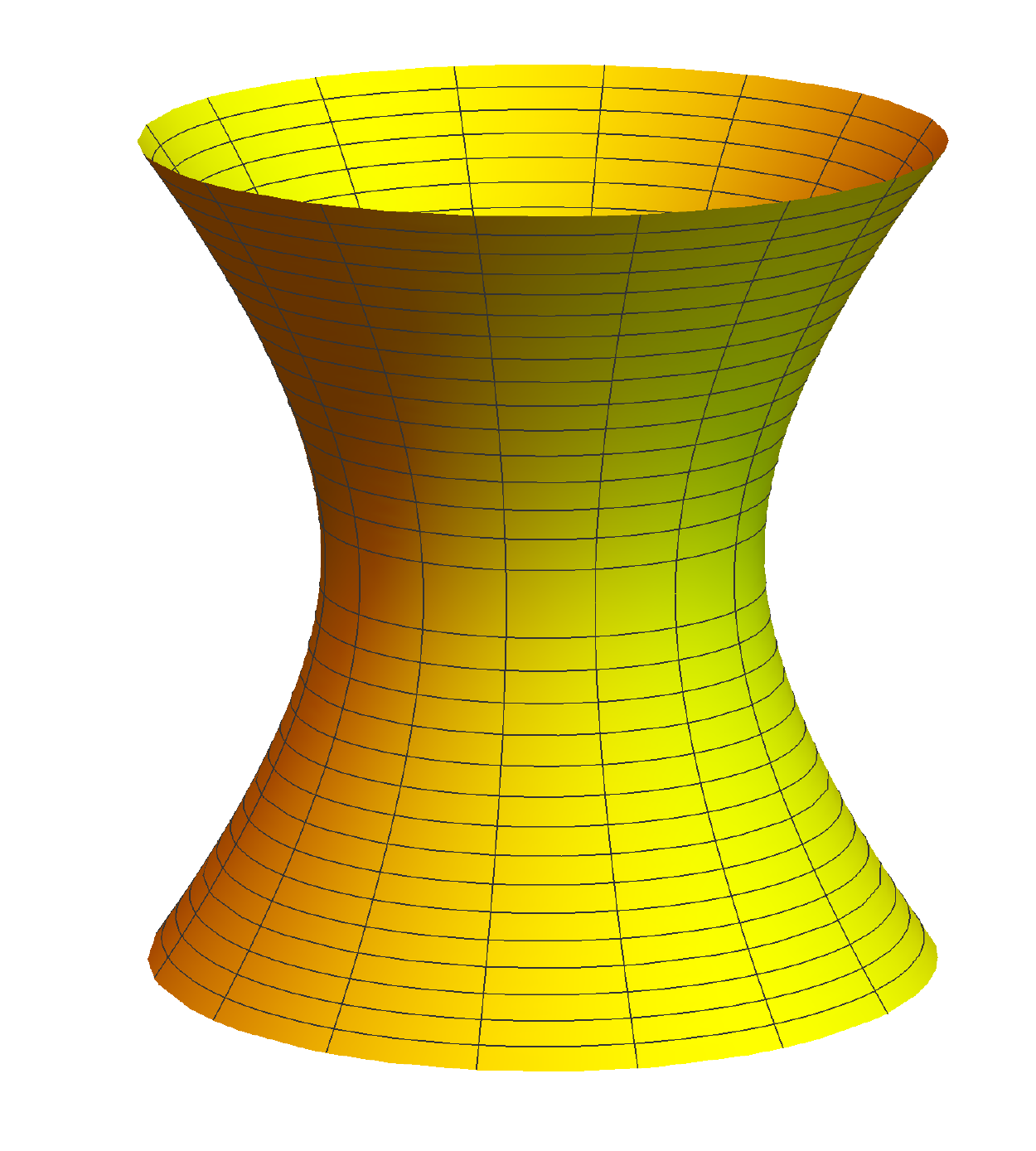}
(b)\includegraphics[scale=0.35]{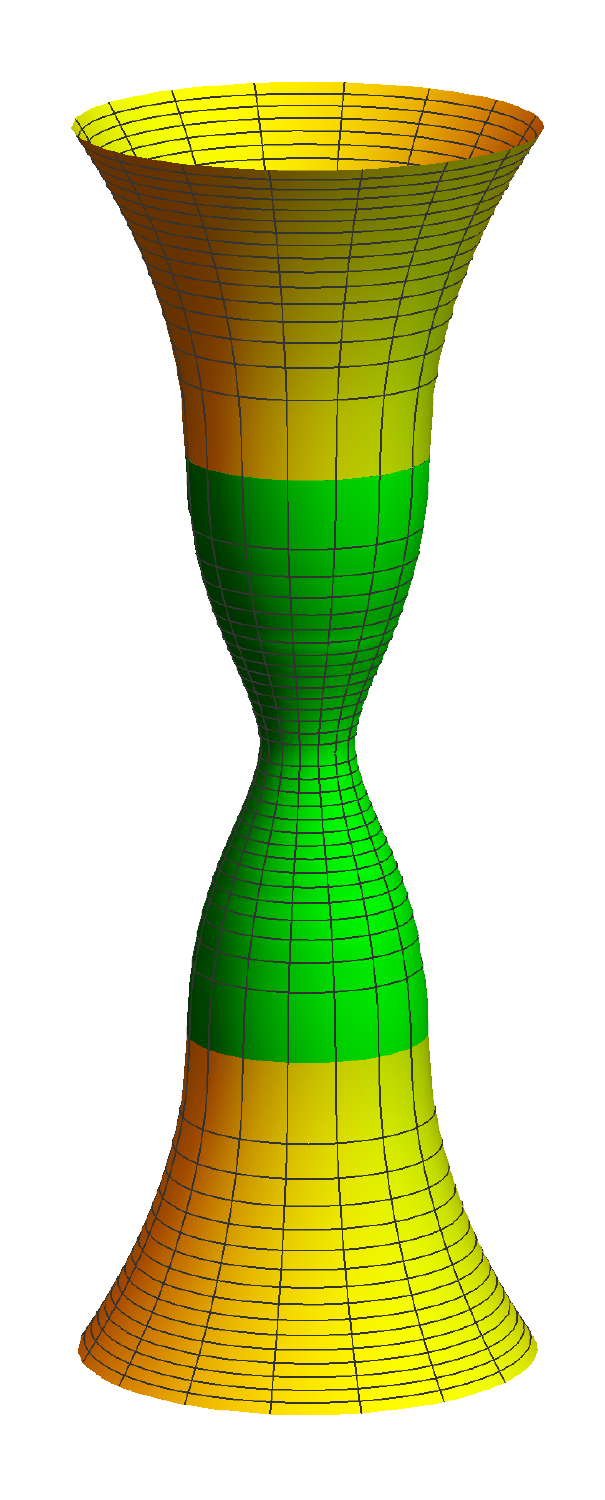}
(c)\includegraphics[scale=0.35]{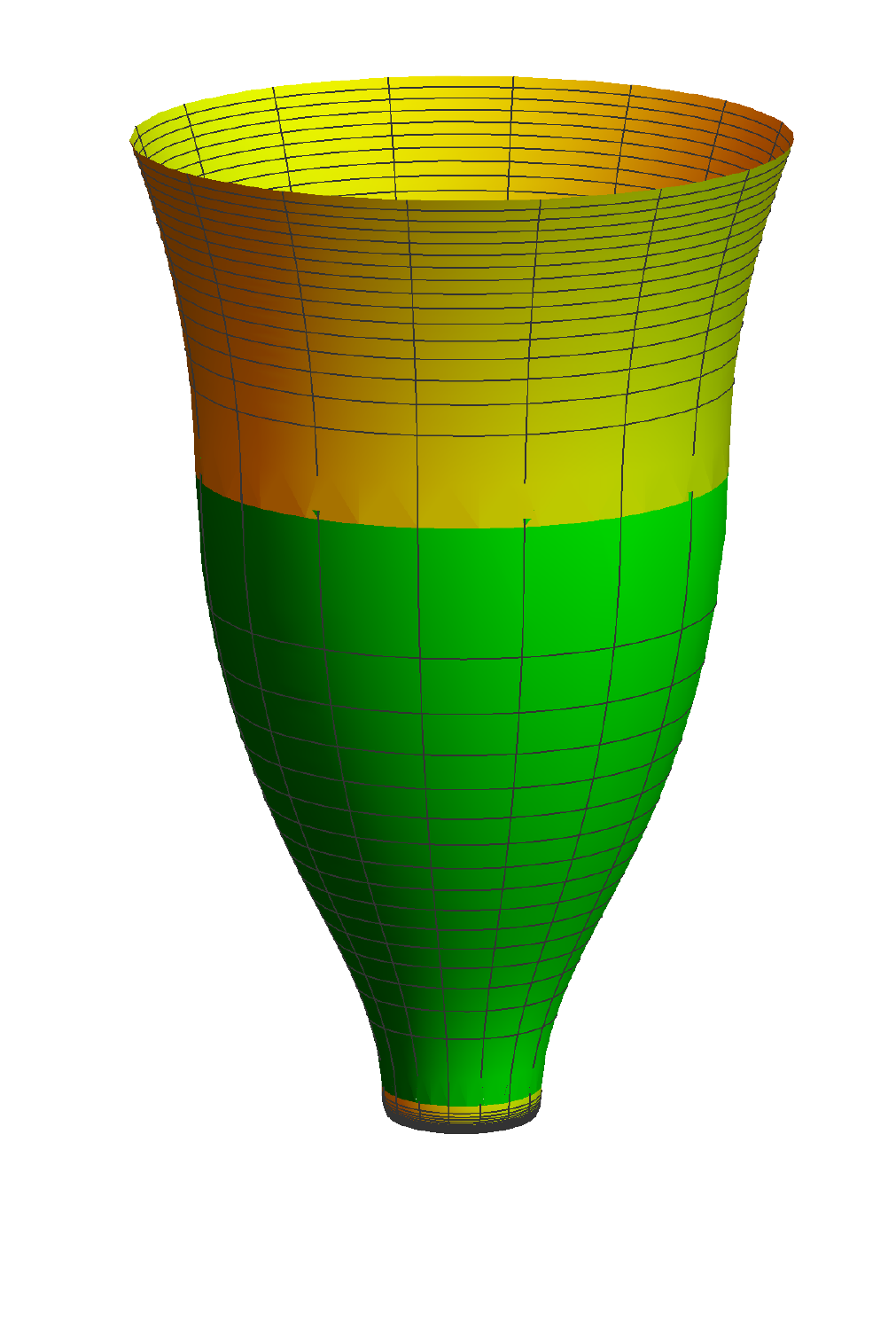}
\caption{\label{fig.emb}The embedding diagram of rLQGO. (a): A timelike wormhole without horizon (region I). (b): A spacelike transition surface inside the event horizon (region II). (c): The transition surface is inside the inner horizon (region III).}
\end{figure}

\begin{figure}
\center(a)\includegraphics[scale=0.3]{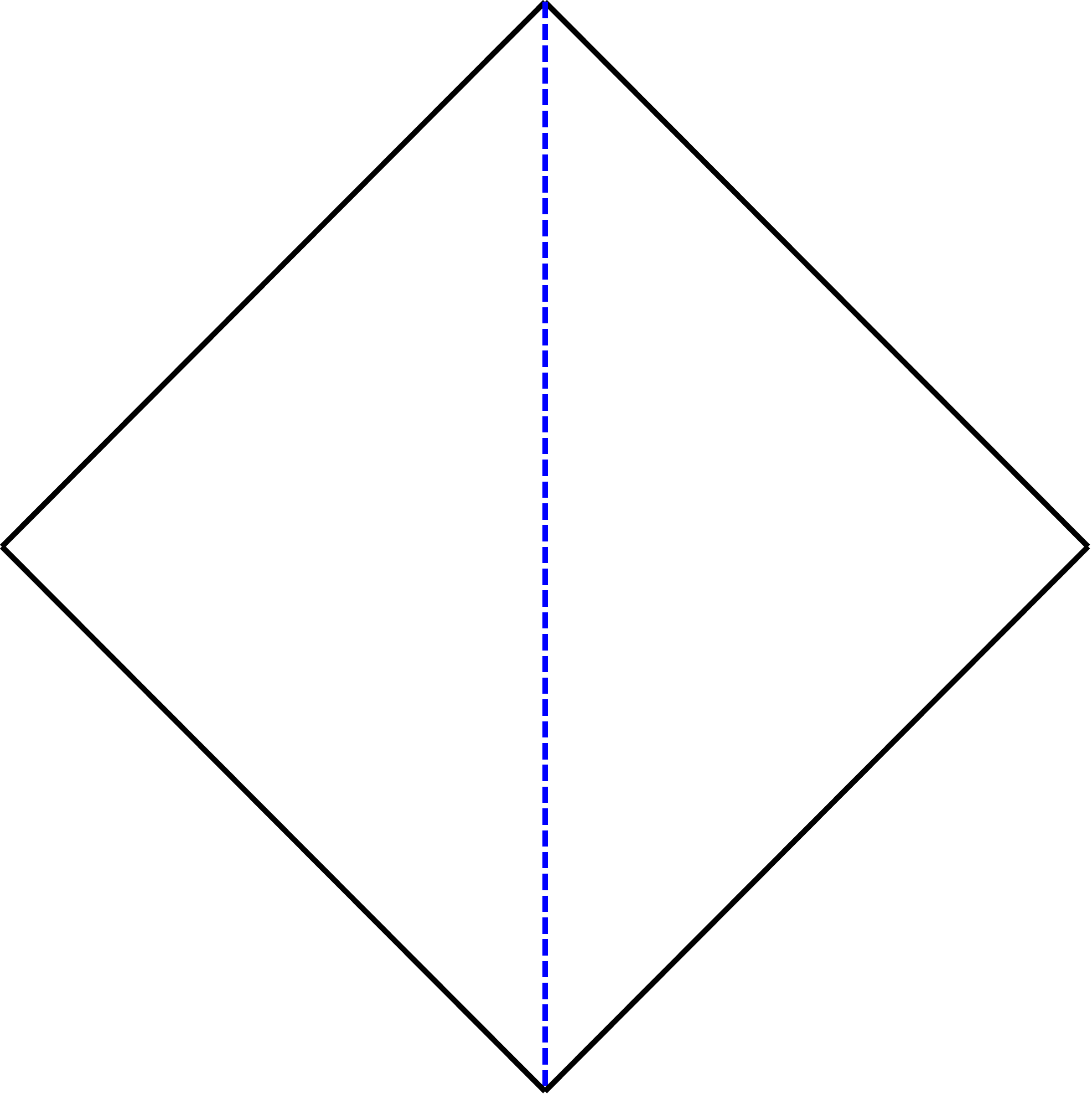}
(b)\includegraphics[scale=0.2]{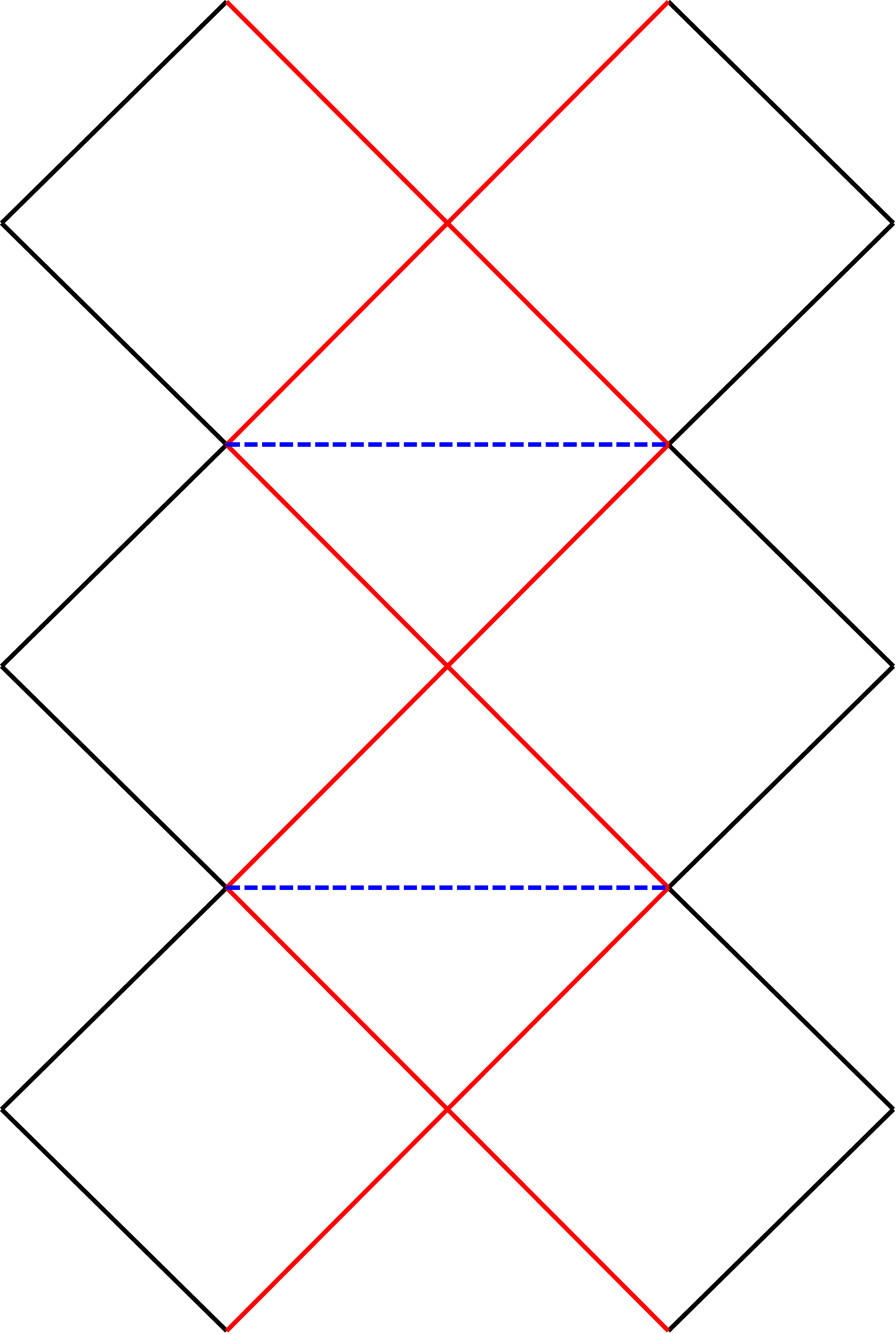}
\caption{\label{fig.penrose1}The Penrose diagram of rLQGO in (a): region I and (b): region II. Blue dashed lines denote the transition surface and slanted red lines, at an angle of $45^{\circ}$, are event horizons.}
\end{figure}

\begin{figure}
\center\includegraphics[scale=0.18]{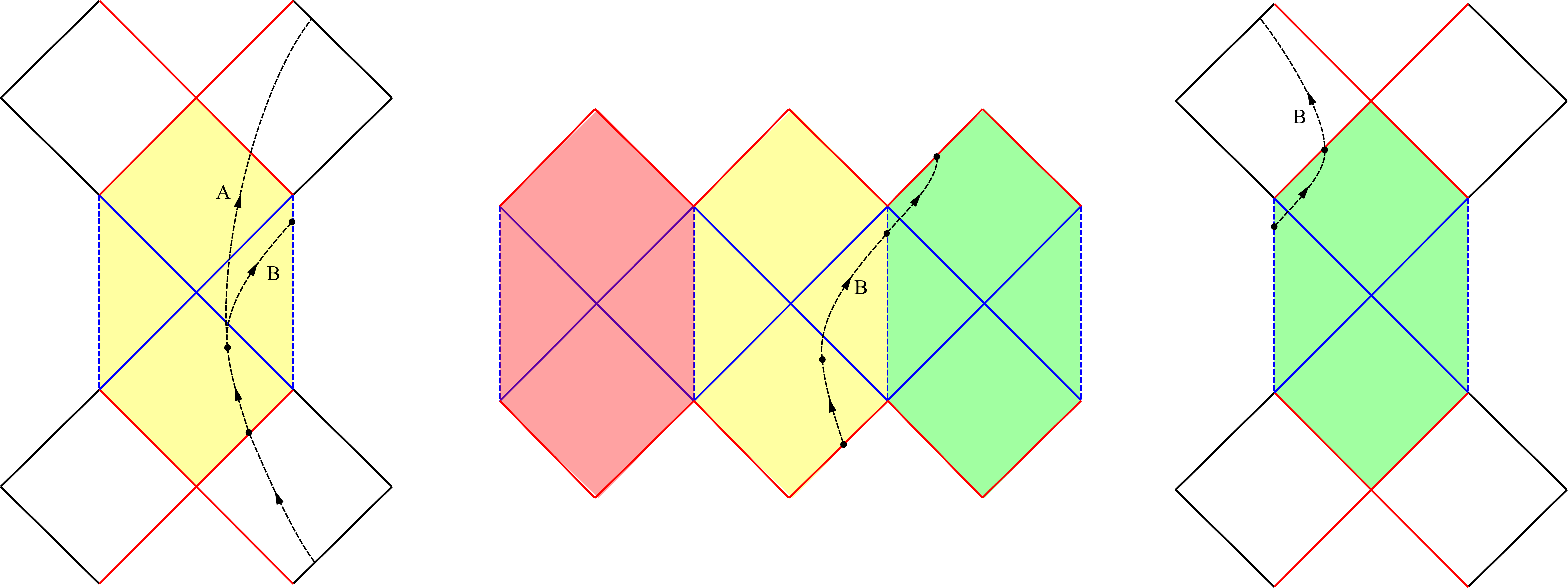}
\caption{\label{fig.penrose3}The Penrose diagram of rLQGO in region III. The colored regions are inside the outer event horizon (red lines). After entering inner horizons (solid blue lines), a timelike trajectory can be extended to another universe either by going upward (trajectory A) without touching the transition surface (dashed blue lines), or by crossing the transition surface to another interior patch (trajectory B). Note that the exterior regions of the two adjacent interior patches can be causally disconnected.}
\end{figure}

Importantly, the rLQGO is free from spacetime singularities. As shown in Figure~\ref{fig.Ricci}, the Ricci scalar is finite everywhere on the $(y,\cos\theta)$ plane, and rapidly vanishes when moving away from the transition surface. In this figure, the solid red and blue lines represent the outer and inner horizons, respectively. The dashed curves are the ergosurface. Interestingly, since the areal radius $b\neq0$ for rLQGO, there is no closed timelike curve which usually appears near the ring singularity inside the Kerr black hole (see the Appendix for details). Even though the event horizon disappear when $|a|>M_B$, the absence of singularity naturally preserves the Cosmic censorship hypothesis and the configuration looks like a ``naked bounce'' (similar to the rLQGO solution in Region I). 

\begin{figure*}[t]
\center\includegraphics[scale=0.35]{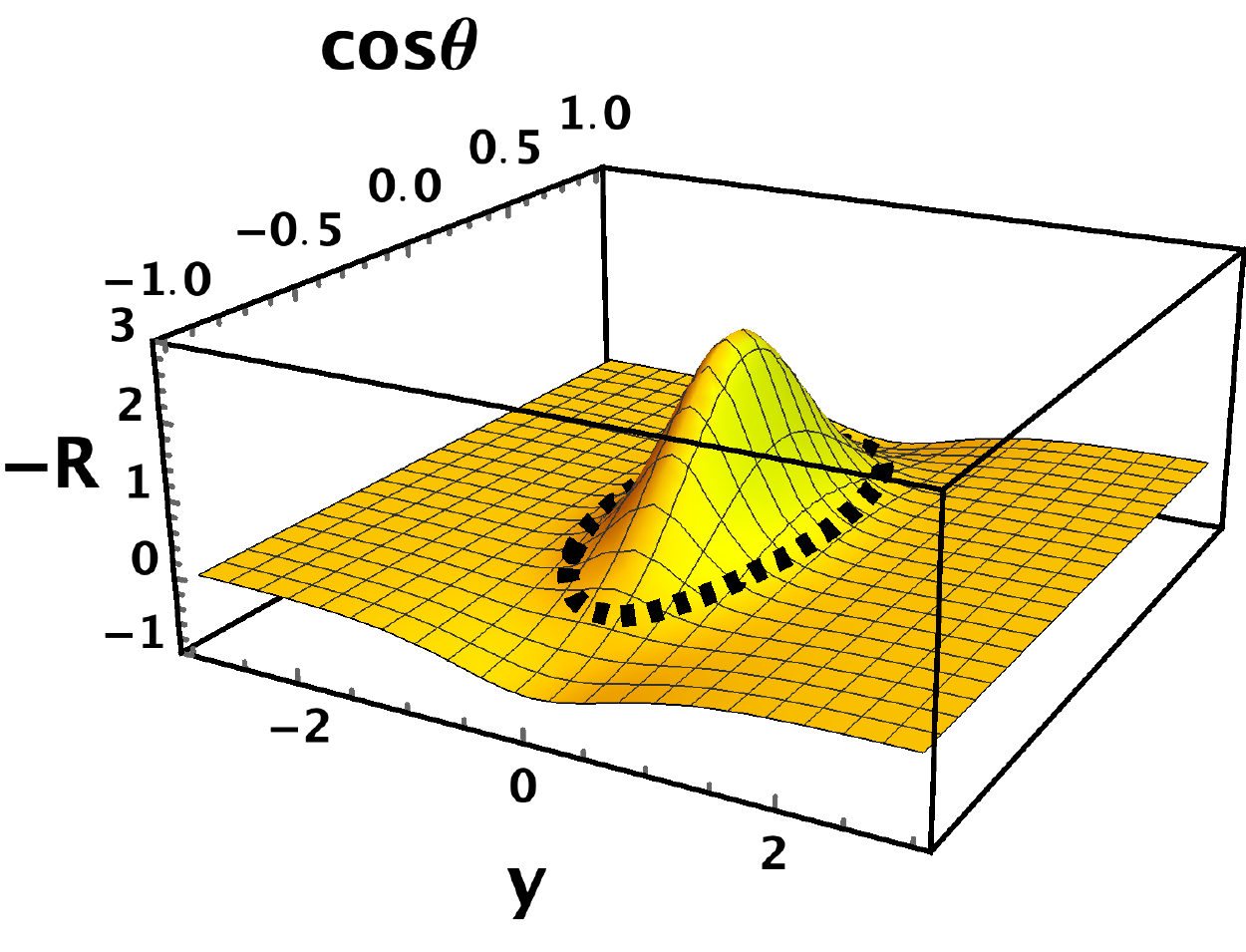}(a)
\includegraphics[scale=0.35]{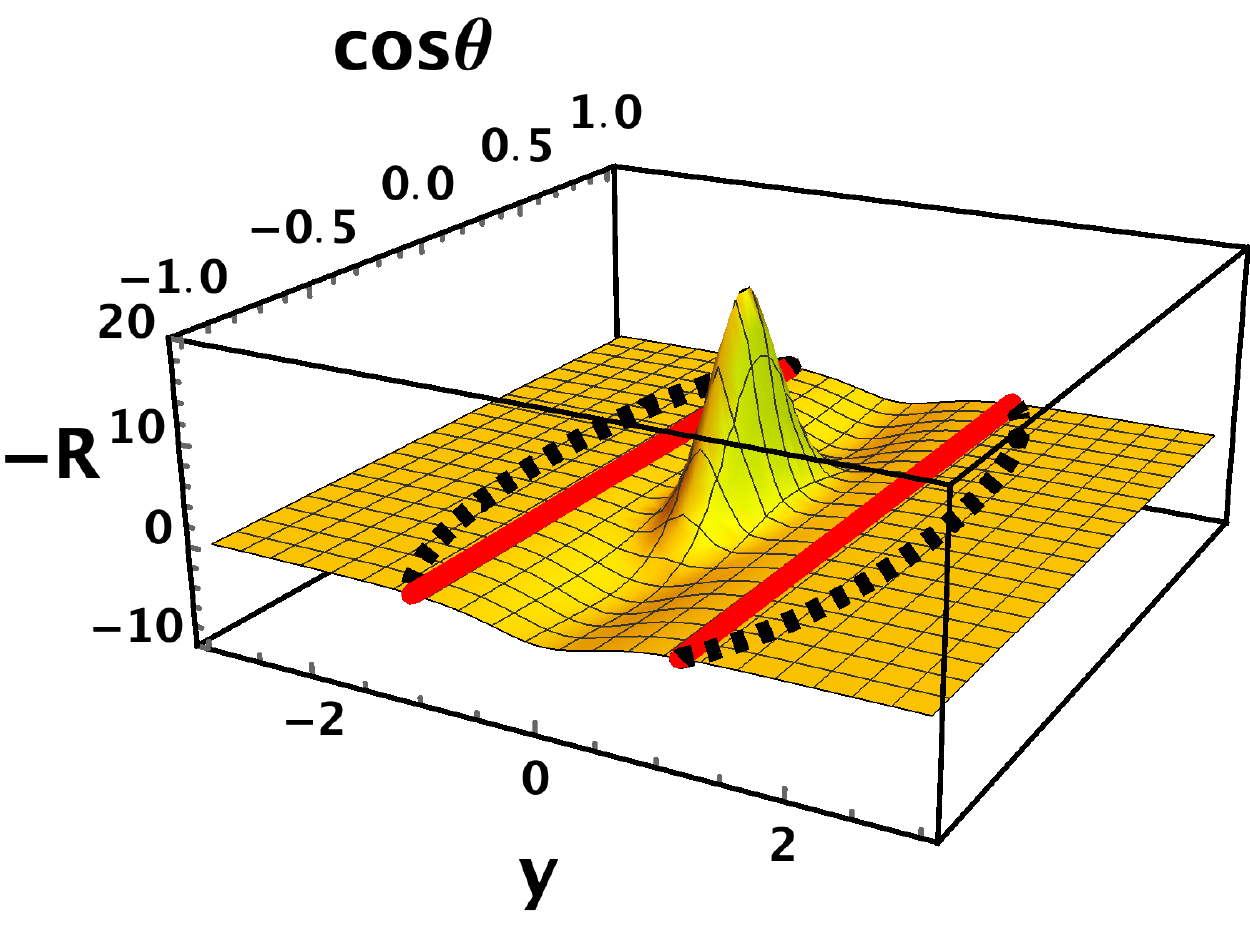}(b)
\includegraphics[scale=0.35]{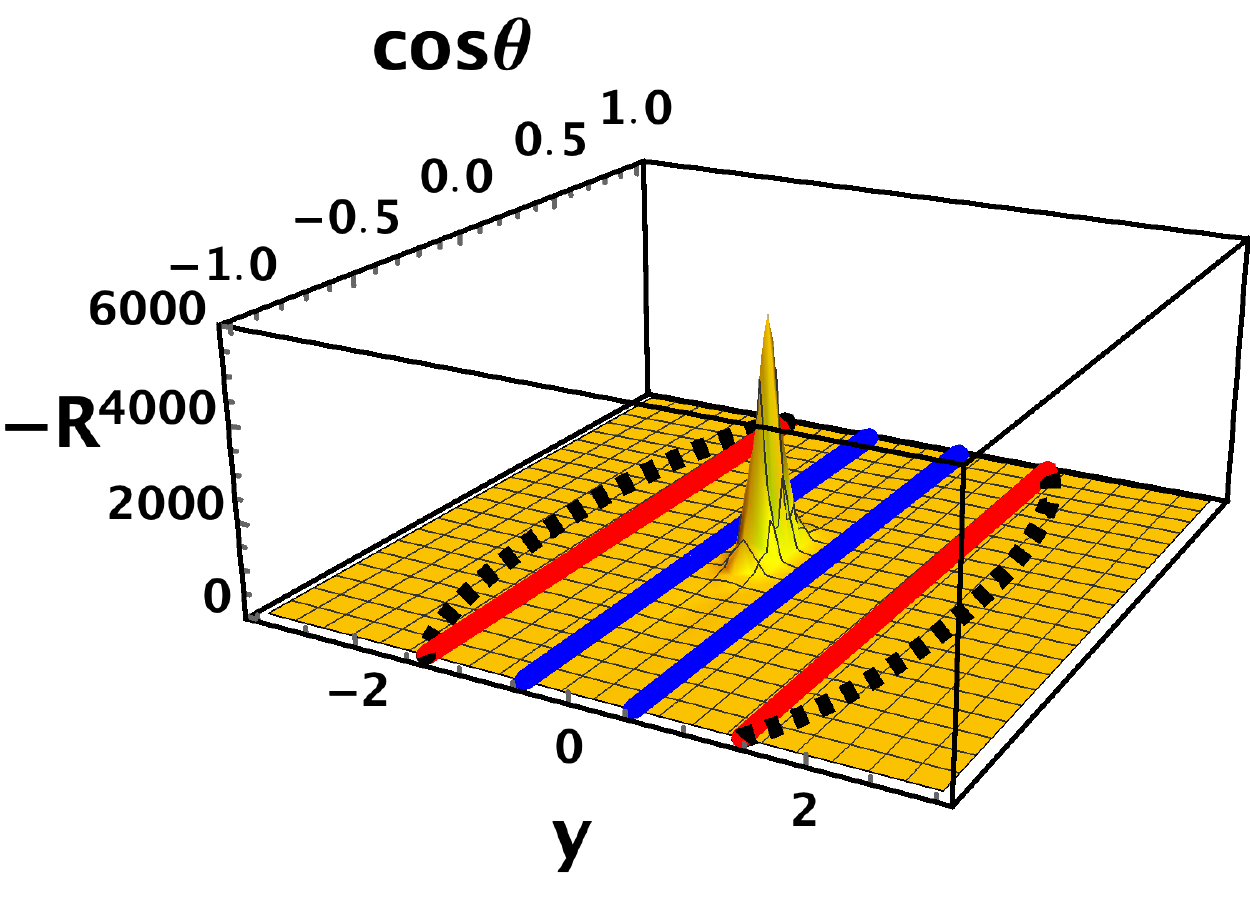}(c)
\caption{\label{fig.Ricci}The Ricci scalar $R$ of the rLQGO spacetime expressed in the $(y,\cos\theta)$ plane. In the figure, we set $M_B=1$ and $a/M_B=0.9$. The solid red and blue lines represent the outer and the inner event horizons. The dashed curves represent the ergosurface. (a): A wormhole without horizon (region I with $A_\lambda=0.4$). (b): The transition surface is covered by the outer horizon (region II with $A_\lambda=0.1$) (c): The transition surface is inside the inner horizon (region III with $A_\lambda=0.01$).}
\end{figure*}

\section{Astrophysical implications}
In addition to having properties of, \textit{e.g.} asymptotic flatness and regularity, we find that both the geodesic equations and the Klein-Gordon equation of the rLQGO allow for a complete separation of variables (see the Appendix), following the criteria of Refs.~\cite{Shaikh:2019fpu,Chen:2019jbs}. The separability of the geodesic equations is useful in testing the rLQGO spacetime with its shadow and the orbital motion of surrounding particles while the separability of the Klein-Gordon equation is helpful for studying the scattering problem and the quasinormal modes \cite{prepare}.

As an example, let us demonstrate that, it is possible, in principle, to constrain the quantum parameter $A_\lambda$ using the shadow image cast by the M87*. In particular, we find that the effects made by the parameter $A_\lambda$ on the shadow size $R_S/M_B$ are more significant than those on the non-circularity of the shadow contour. Provided that the shadow size cast by M87* is consistent with that of Kerr black hole within $17\%$ at $1\sigma$ level \cite{Akiyama:2019eap,Psaltis:2020lvx}, one can see from Figure~\ref{fig.shadowRS} that the parameter space corresponding to the wormhole geometry (the region on the right of the red curve) is disfavored by the bound from the $R_S$ measurement (black curve). Since the quantum parameter which enters the effective equations in LQG is directly related to the fundamental \textit{area gap} in the theory, shadows of rotating black holes give us a new way to constrain this parameter from observations. (Note that the quantum parameter is more tightly constrained by Solar System tests,  $A_\lambda < 7.7\times 10^{-5}$ \cite{Williams:2004qba}. However, this assumes the validity of Birkhoff's theorem which need not hold in LQG.) This paves a novel method for deriving state-of-the-art constraints on LQG by examining observational consequences of rLQGO \cite{prepare}.

\begin{figure}[t]
\center\includegraphics[scale=0.71]{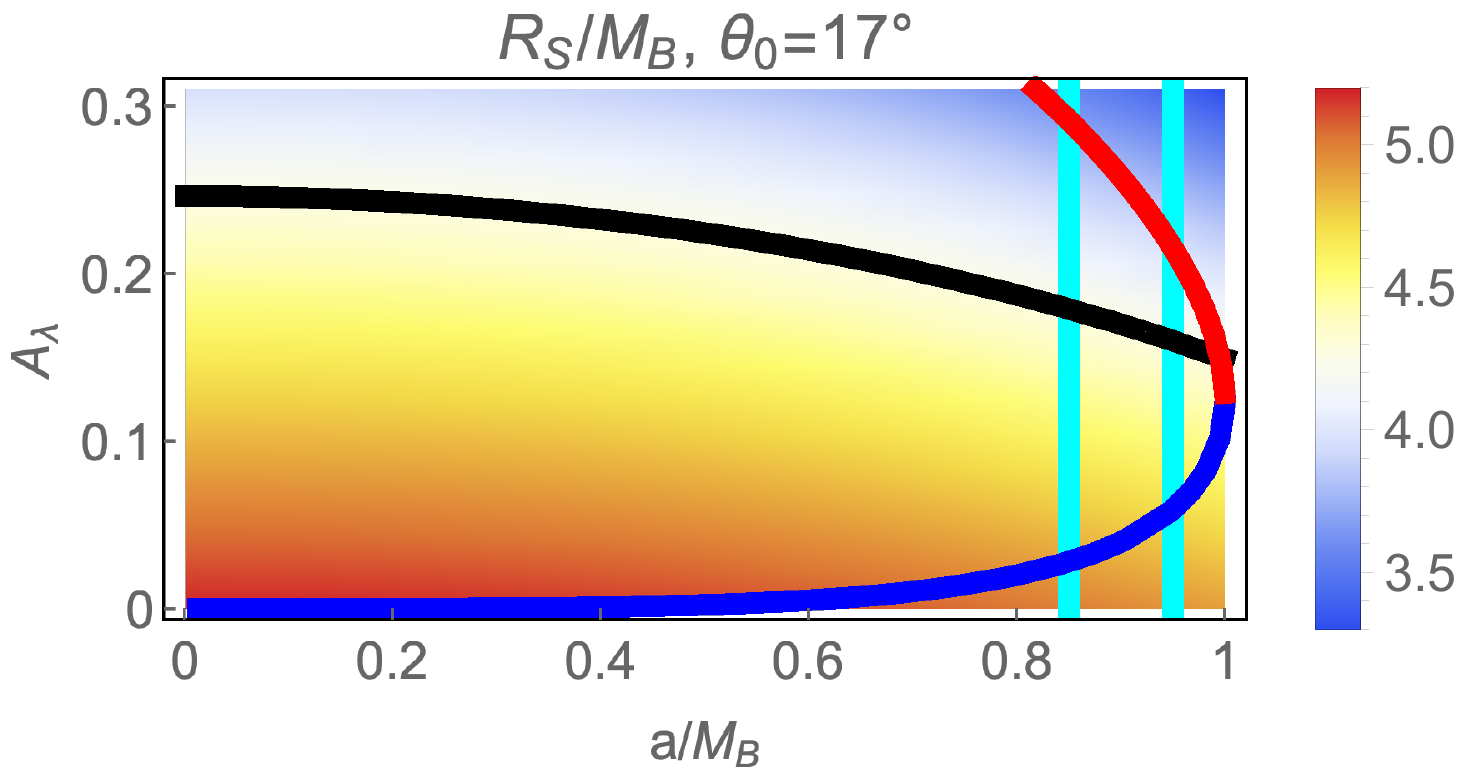}
\caption{\label{fig.shadowRS}The apparent size $R_S/M_B$ of the shadow cast by rLQGO is shown with respect to the parameter space $\{a,A_\lambda\}$. The $17\%$ bound of $R_S/M_B$ (black curve) inferred from the M87* shadow disfavors parameters corresponding to the wormhole geometry (the region to the right of the red curve). Here we have taken into account the spin measurement obtained using the radio intensity data \cite{Tamburini:2019vrf} (cyan lines), and the inclination angle measured by the jet direction \cite{Walker:2018muw}.}
\end{figure}

\section{Universal features}
The most obvious limitation of our approach is that the resulting rLQGO metric is not derived by a direct loop quantization of the Kerr (or, more generally, axisymmetric) spacetime. How much of our results do we expect to generalize to such a scenario, and not be tied to the seed metric that we have chosen? Firstly, note that the existence of a \textit{spacelike} transition surface is very common for non-rotating LQGBHs, irrespective of quantization ambiguities (such as choosing the $\mu_0$ or the $\bar\mu$ scheme). Since this is the most crucial feature of the seed metric we have used in our construction, it is natural to expect that our rLQGO solution correctly captures the effective spacetime description of rotating LQGBHs, as long as we expect LQG to provide singularity-resolution of rotating black holes in a way such that there is a smooth bouncing geometry bridging black and white holes. Furthermore, our results indicate that such a geometry observationally favors having the transition surface inside the inner horizon, and is automatically consistent with the expectation that the quantum parameter is small (it inherits this property from the tiny area gap $\propto\ell_{\rm Pl}^2$). Interestingly, observations also seem to rule out models of non-rotating LQGBH spacetimes which describe a bounce \textit{outside} the event horizon \cite{Haggard:2014rza}, since their rotating counterparts are at odds with observations, as well as prefer non-rotating models which allow for an inner horizon \cite{LQG2, LQG4}.

To make our point more explicit, we present the result of the NJA analyses on another non-rotating LQG metric proposed in \cite{LQG2,LQG3} (see the Appendix and \cite{prepare} for details). In Figure~\ref{fig.edshadow}, we show the apparent size $R_S/M_B$ of the shadow cast by the rotating metric, which is obtained from \cite{LQG2,LQG3} using NJA, in the parameter space of $\{a,\tilde{\Delta}\}$. Firstly, we note that the spin and the quantum parameter $\tilde{\Delta}$ both shrink the shadow size, exactly similar to the rLQGO case. Secondly, the black boundary represents a spin-dependent upper bound of the quantum parameter, above which the object cannot cast shadows. In the non-rotating limit, this upper bound can be explicitly derived as $\gamma^2\tilde{\Delta}/M_B^2<3^6/2^{10}\approx0.71$, where $\gamma$ is the Immirzi parameter and $\tilde{\Delta}$ is the area-gap in LQG which is directly constrained in this model. Thus, we find other models of regular LQGBHs also support our general finding that the area-gap is constrained to be small from observations.

\begin{figure}[h]
	\center\includegraphics[scale=0.71]{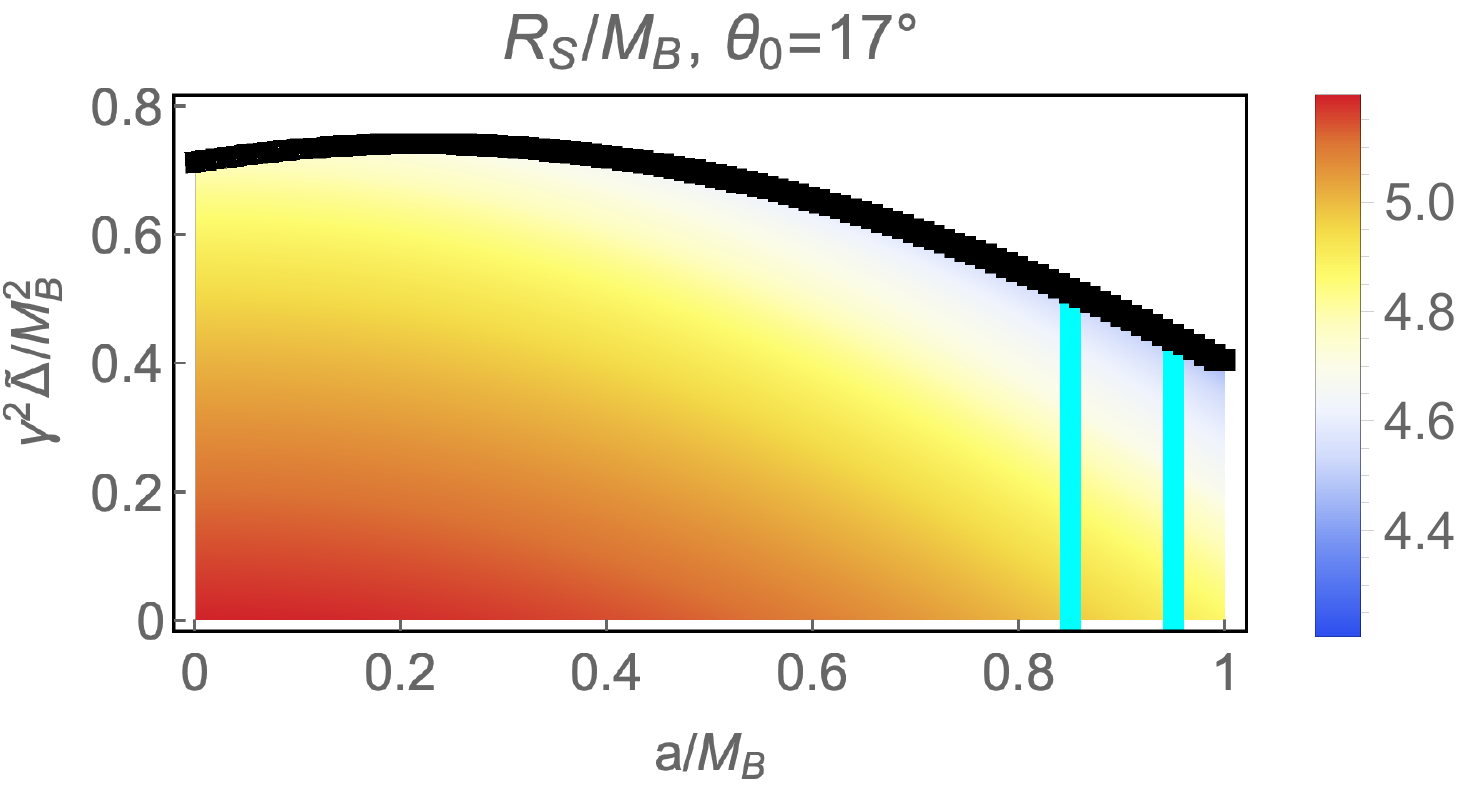}
	\caption{\label{fig.edshadow}The apparent size $R_S/M_B$ of the shadow cast by the rotating black hole metric, corresponding to \cite{LQG2,LQG3}, in the parameter space of $\{a,\tilde{\Delta}\}$. The black boundary represents a spin-dependent upper bound of the quantum parameter, above which shadow contours disappear.}
\end{figure}

\section{Discussions}
The construction of rotating LQGBHs from holonomy-corrected effective equations in LQG is still an open problem. In order to catch up with the rapidly developing astronomical observations of spinning black holes in the coming years, there is an urgent need for a model of rotating LQGBH. To derive this, an alternate path is to use a viable solution-generating method to generate a rotating solution from a non-rotating LQGBH seed metric. 
The resulting rLQGO spacetime \eqref{finalrot}, based on the seed metric \cite{Bodendorfer:2019jay, Bodendorfer:2019nvy}, possesses a rather simple expression and has several interesting properties. It is everywhere non-singular and it reduces to Kerr solution asymptotically. The geodesic equations and the Klein-Gordon equation both allow complete separations of variables. Most importantly, as in the static LQGBH, the rLQGO is characterized by the existence of a transition surface induced from non-perturbative quantum corrections. Depending on the relative location of the transition surface with respect to the two horizons, the rLQGO can represent a wormhole, a regular black hole with an interior spacelike transition surface, or a regular black hole with a timelike transition region inside the inner horizon. We show that the possibility of rLQGO being a wormhole without horizon has been almost \textit{ruled out} by the shadow size of M87* measured by EHT.

Most significantly, our works fills a lacuna between theoretical quantum gravity extensions of black holes, which have been mostly applied to non-rotating spacetimes, and experimental observations which have been of spinning black holes. Remarkably, we find that not only is it possible to find a regular effective description of rotating black holes from LQG, an extension to such backgrounds leads to observable effects which can rule out some proposals of loop quantization for non-rotating black holes while providing support for other more generic ones which capture some universal features of LQG.

\section*{Acknowledgements} SB is grateful to Jibril Ben Achour, Norbert Bodendorfer and Johannes M\"unch for comments on an earlier version of this draft. SB is supported in part by the NSERC (funding reference \#CITA 490888-16) through a CITA National Fellowship and by a McGill Space Institute fellowship. CYC is supported by Institute of Physics in Academia Sinica, Taiwan. DY is supported by the National Research Foundation of Korea (Grant no. 2018R1D1A1B07049126).

\appendix
\section*{Appendix}
In this Appendix, we will present the detailed derivation of the metric of the rotating Loop Quantum Gravity compact object (rLQGO). The derivation is based on the Newman-Janis Algorithm (NJA) \cite{Newman:1965tw} in which one starts with a non-rotating Loop Quantum Gravity black hole (LQGBH) as the seed metric, then introduces the spin by performing a complex shift on the advanced null coordinates. The LQGBH seed metric that we are going to consider is
\begin{equation}
	ds^2=-8A_\lambda M_B^2\tilde{a}(y)dt^2+\frac{dy^2}{8A_\lambda M_B^2\tilde{a}(y)}+b(y)^2d\Omega_2^2\,,\label{BMMmetricformal}
\end{equation}
where \cite{Bodendorfer:2019nvy,Bodendorfer:2019jay}
\begin{align}
	b(x)^2=&\,\frac{A_\lambda}{\sqrt{1+x^2}}\frac{M_B^2\left(x+\sqrt{1+x^2}\right)^6+M_W^2}{\left(x+\sqrt{1+x^2}\right)^3}\,,\label{BMM2metrica1A}\\
	\tilde{a}(x)=&\,\left(1-\sqrt{\frac{1}{2A_\lambda}}\frac{1}{\sqrt{1+x^2}}\right)\frac{1+x^2}{b(x)^2}\,,\label{BMM2metricfunctionsA}
\end{align}
and $y\equiv\sqrt{8A_\lambda}M_Bx$. We will focus on the symmetric bouncing model in which $M_B=M_W$. As will be illustrated later, we will in particular adopt the revised version of NJA \cite{Azreg-Ainou:2014pra} to recast the resultant metric in the Boyer-Lindquist coordinate system. After obtaining the rotating metric (Sec.~\ref{sec.NJA}), we will present its asymptotic expression and the Arnowitt-Deser-Misner (ADM) mass. Then, we will show that there is no closed timelike curve in rLQGO, as opposed to the classical Kerr spacetime (Sec.~\ref{sec.ccc}). After that, we will explicitly show that the geodesic equations (Sec.~\ref{sec.geo}) and the Klein-Gordon (KG) equation (Sec.~\ref{sec.KG}) in the rLQGO spacetime allow complete separations of variables.

\section{NJA and rLQGO metric}\label{sec.NJA}
We consider a general static and spherically symmetric seed metric:
\begin{equation}
	ds^2=-g(y)dt^2+\frac{dy^2}{f(y)}+b(y)^2d\Omega_2^2\,.\label{seedmetricNJA}
\end{equation}
The first step of NJA is to introduce a set of advanced null coordinates ($u,y,\theta,\phi$), where
\begin{equation}
	u\equiv t-y_*\,,\qquad \frac{dy_*}{dy}\equiv\frac{1}{\sqrt{fg}}\,.
\end{equation}
In the advanced null coordinates, one can express the inverse of the seed metric by using a null tetrad $Z_a^\mu=(l^\mu,n^\mu,m^\mu,\bar{m}^\mu)$ as follows
\begin{equation}
	g^{\mu\nu}=-l^\mu n^\nu- l^\nu n^\mu +m^\mu \bar{m}^\nu +m^\nu \bar{m}^\mu\,,\label{ginversetetrad}
\end{equation}
where $\bar{m}^\mu$ is the complex conjugate of $m^\mu$. In the advanced null coordinates, the null tetrad can be expressed as
\begin{align}
	l^\mu&=\delta_y^\mu\,, \qquad n^\mu=\sqrt{\frac{f(y)}{g(y)}}\delta_u^\mu-\frac{f(y)}{2}\delta_y^\mu\,,\nonumber\\
	m^\mu&=\frac{1}{\sqrt{2}b(y)}\left(\delta_\theta^\mu+\frac{i}{\sin\theta}\delta_\phi^\mu\right)\,.
\end{align}
In NJA, the key step to include spin is by performing a complex shift on the advanced null coordinates
\begin{equation}
	u'=u-ia\cos\theta\,,\qquad y'=y+ia\cos\theta\,,\label{complexshi}
\end{equation}
where $a$ is regarded as the spin of the spacetime. The angular coordinates $\theta$ and $\phi$ remain intact. It should be mentioned that after the complex shift \eqref{complexshi}, the coordinates $(u',y',\theta,\phi)$ are required to be real-valued. 

In the standard NJA, the metric functions after the complex shift would generically be functions of $y$ and $\theta$. The explicit expression of the metric functions depends on how one invokes the complexification procedure on the radial coordinate $y$, hence inevitably contains some arbitrariness. However, in the revised version of NJA \cite{Azreg-Ainou:2014pra}, the complexification procedure is replaced with another criterion. For the time being, we denote the metric functions after the complex shift by
\begin{equation}
	f(y)\rightarrow F(y,\theta)\,,\quad g(y)\rightarrow G(y,\theta)\,,\quad b(y)^2\rightarrow \Psi(y,\theta)\,,
\end{equation}  
where we have dropped the prime for the sake of simplicity. 

After the complex shift, one can obtain a new set of null tetrad basis from which, by using Eq.~\eqref{ginversetetrad}, the corresponding new line element in the advanced null coordinates can be written as
\begin{align}
	ds^2=&-2\sqrt{\frac{G}{F}}dudy+2a\sin^2\theta\left(G-\sqrt{\frac{G}{F}}\right)dud\phi\nonumber\\
	&-Gdu^2+\Psi d\theta^2+2a\sqrt{\frac{G}{F}}\sin^2\theta dyd\phi\nonumber\\
	&+\sin^2\theta\left[\Psi+a^2\sin^2\theta\left(2\sqrt{\frac{G}{F}}-G\right)\right]d\phi^2\,.\label{NJAu}
\end{align}

The last step of NJA is to rewrite the metric \eqref{NJAu} in the Boyer-Lindquist coordinate system $(t,y,\theta,\varphi)$, such that the $g_{t\varphi}$ component is the only off-diagonal component. This can be achieved by adopting the following coordinate transformations:
\begin{equation}
	du=dt+\lambda_1(y)dy\,,\qquad d\phi=d\varphi+\lambda_2(y)dy\,,\label{trans}
\end{equation}
where 
\begin{align}
	\lambda_1(y)=-\frac{\sqrt{\frac{F(y,\theta)}{G(y,\theta)}}\Psi(y,\theta)+a^2\sin^2\theta}{F(y,\theta)\Psi(y,\theta)+a^2\sin^2\theta}\,,\nonumber\\
	\lambda_2(y)=-\frac{a}{F(y,\theta)\Psi(y,\theta)+a^2\sin^2\theta}\,.
\end{align}
It should be emphasized that $\lambda_1$ and $\lambda_2$ should be functions of $y$ only, otherwise the transformations \eqref{trans} would be invalid. The problem associated with the standard complexification procedure in NJA for determining ($F,G,\Psi$) is that these metric functions in general do not guarantee the validity of the coordinate transformations \eqref{trans}. However, in the revised version of NJA \cite{Azreg-Ainou:2014pra}, the complexification procedure is skipped and the metric functions $F$ and $G$ are chosen exquisitely such that the transformations \eqref{trans} are ensured to be well-defined. In the revised NJA, the metric functions are assumed to be
\begin{align}
	G(y,\theta)&=\frac{f(y)b(y)^2+a^2\cos^2\theta}{\left(\sqrt{\frac{f(y)}{g(y)}}b(y)^2+a^2\cos^2\theta\right)^2}\Psi(y,\theta)\,,\\
	F(y,\theta)&=\frac{f(y)b(y)^2+a^2\cos^2\theta}{\Psi(y,\theta)}\,,
\end{align}
with which one can easily verify that
\begin{align}
	\lambda_1(y)&=-\frac{\sqrt{\frac{f(y)}{g(y)}}b(y)^2+a^2}{f(y)b(y)^2+a^2}\,,\nonumber\\
	\lambda_2(y)&=-\frac{a}{f(y)b(y)^2+a^2}\,.
\end{align}
Therefore, the transformations \eqref{trans} are guaranteed to be valid. As a result, the metric of the rotating spacetime in the Boyer-Lindquist coordinate can be written as
\begin{align}
	ds^2=&\,\Psi\Bigg\{-\frac{fb^2+a^2\cos^2\theta}{\left(K+a^2\cos^2\theta\right)^2}dt^2+d\theta^2+\frac{dy^2}{fb^2+a^2}\nonumber\\
	&+2a\sin^2\theta\left[\frac{fb^2-K}{\left(K+a^2\cos^2\theta\right)^2}\right]dtd\varphi\nonumber\\
	+\sin^2&\theta\left[1+\frac{a^2\sin^2\theta\left(2K+a^2\cos^2\theta-fb^2\right)}{\left(K+a^2\cos^2\theta\right)^2}\right]d\varphi^2\Bigg\}\,,\label{rotatingintermin123}
\end{align}
where $K=K(y)\equiv b^2\sqrt{f/g}$.

Adopting the following redefinitions \cite{Azreg-Ainou:2014aqa}: $\rho^2\equiv K+a^2\cos^2\theta$, $M=M(y)\equiv\left(K-fb^2\right)/2b$, $\Delta\equiv fb^2+a^2$, and $\Sigma\equiv\left(K+a^2\right)^2-a^2\Delta\sin^2\theta$, the rotating metric \eqref{rotatingintermin123} can be written in a Kerr-like form:
\begin{align}
	ds^2=\frac{\Psi}{\rho^2}&\Bigg[-\left(1-\frac{2Mb}{\rho^2}\right)dt^2-\frac{4aMb\sin^2\theta}{\rho^2}dtd\varphi\nonumber\\&+\rho^2d\theta^2+\frac{\rho^2dy^2}{\Delta}+\frac{\Sigma\sin^2\theta}{\rho^2}d\varphi^2\Bigg]\,.\label{psirhometric}
\end{align}
Note that at this point, the metric function $\Psi(y,\theta)$ remains undetermined. 

By comparing the non-rotating LQGBH metric \eqref{BMMmetricformal} and the seed metric of NJA \eqref{seedmetricNJA}, one identifies
\begin{equation}
	f(y)=g(y)=8A_\lambda M_B^2\tilde{a}(y)\,,
\end{equation}
which implies $K=b^2$. Requiring the rotating metric to recover the static one \eqref{BMMmetricformal} when $a\rightarrow0$, it is natural to set the metric function $\Psi$ to be
\begin{equation}
	\Psi(y,\theta)=\rho^2=b^2+a^2\cos^2\theta\,.\label{psichoise}
\end{equation}
Although the rotating spacetime with the choice of $\Psi$ given by Eq.~\eqref{psichoise} may not satisfy the diagonal components of the Einstein equations \cite{Azreg-Ainou:2014aqa} (it dose at the asymptotic limit), this choice seems to be the most natural one from physical points of view because it naturally satisfies the requirement that the rotating spacetime should recover the static one when $a\rightarrow0$. In addition, from a phenomenological perspective, the Einstein equations are not expected to be satisfied when LQG effects become significant because the spacetime is not classical anymore. Another interesting advantage of choosing \eqref{psichoise} is that, as will be shown later, the geodesic equations as well as the KG equation allow complete separations of variables. In fact, since the metric function $\Psi$ enters the line element \eqref{psirhometric} in the form of a conformal factor, it does not contribute to the photon geodesic equations at all \cite{Shaikh:2019fpu,Junior:2020lya}.

As a summary, the metric of the rLQGO reads
\begin{align}
	ds^2=&-\left(1-\frac{2Mb}{\rho^2}\right)dt^2-\frac{4aMb\sin^2\theta}{\rho^2}dtd\varphi\nonumber\\&+\rho^2d\theta^2+\frac{\rho^2dy^2}{\Delta}+\frac{\Sigma\sin^2\theta}{\rho^2}d\varphi^2\,,\label{finalrotapp}
\end{align}
where
\begin{align}
	\rho^2&=b^2+a^2\cos^2\theta\,,\qquad M=\frac{b\left(1-8A_\lambda M_B^2\tilde{a}\right)}{2}\,,\nonumber\\
	\Delta&=8A_\lambda M_B^2\tilde{a}b^2+a^2\,,\qquad \Sigma=\left(b^2+a^2\right)^2-a^2\Delta\sin^2\theta\,.
\end{align}

At the asymptotic region where $|y|\approx b\rightarrow\infty$, the metric components can be approximated as
\begin{align}
	g_{tt}&=-1+\frac{2M_B}{b}-\frac{6A_\lambda M_B^2}{b^2}+\mathcal{O}\left(b^{-3}\right)\,,\nonumber\\ g_{bb}&=1+\frac{2M_B}{r}+\mathcal{O}\left(b^{-2}\right)\,,\nonumber\\
	g_{\theta\theta}&=b^2\left[1+\mathcal{O}\left(b^{-2}\right)\right]\,,\quad g_{\varphi\varphi}=b^2\sin^2\theta\left[1+\mathcal{O}\left(b^{-2}\right)\right]\,,\nonumber\\
	g_{t\varphi}&=-\frac{2M_Ba\sin^2\theta}{b}+\mathcal{O}\left(b^{-2}\right)\,,
\end{align}
Therefore, the rLQGO reduces to Kerr spacetime at the asymptotic region. Furthermore, we can calculate the ADM mass of the rLQGO by firstly considering the $\textrm{constant-}t$ hypersurface $\bold{\Sigma}$ and assuming a zero spin. The induced metric can be written as
\begin{equation}
	ds_\bold{\Sigma}^2=\left(\frac{dy}{db}\right)^2\frac{db^2}{8A_\lambda M_B^2\tilde{a}}+b^2d\Omega_2^2\,.
\end{equation}
Using the induced metric, the ADM mass can be calculated via the following formula \cite{waldbook,Shaikh:2018kfv,Amir:2018pcu}
\begin{align}
	M_{ADM}&=\lim_{b\rightarrow\infty}\frac{b}{2}\left[\left(\frac{dy}{db}\right)^2\frac{1}{8A_\lambda M_B^2\tilde{a}}-1\right]\nonumber\\
	&=M_B\,.
\end{align}
Therefore, the ADM mass of the rLQGO is $M_B$ and it is independent of the quantum parameter $A_\lambda$.

Before ending this section, we would like to mention that in the letter, we also apply NJA to another LQG black hole, which was proposed in \cite{LQG2}, to exhibit that the conclusions drawn in our work are indeed shared among multiple LQG black hole models. Here, we briefly mention the results of this collateral analysis (see \cite{prepare} for more details). 

When written in the Schwarzschild coordinates, the metric of the regular black hole of \cite{LQG2} reads
\begin{align}
ds^2=&-\left(1-\frac{2M_B}{b}+\frac{4\gamma^2\tilde\Delta M_B^2}{b^4}\right)dt^2\nonumber\\&+\left(1-\frac{2M_B}{b}+\frac{4\gamma^2\tilde\Delta M_B^2}{b^4}\right)^{-1}db^2+b^2d\Omega_2^2\,,\label{edmetric}
\end{align}
where $\gamma$ is the Barbero-Immirzi parameter, and $\tilde\Delta$ is the minimal area in LQG. In \cite{LQG2}, it was shown that the radial variable $b$ has again a minimum value due to quantum corrections, and there is no spacetime singularity.

Adopting the NJA onto the metric \eqref{edmetric}, the metric of the rotating generalization is
\begin{align}
ds^2=&-\left(1-\frac{\bar{R}_S(b)b}{\rho^2}\right)dt^2+\rho^2d\theta^2+\frac{\rho^2}{\bar\Delta}db^2\nonumber\\
&-\frac{2a\sin^2\theta\bar{R}_S(b)b}{\rho^2}dtd\varphi+\frac{\bar\Sigma\sin^2\theta}{\rho^2}d\varphi^2\,,\label{edmetricrotating}
\end{align}
where
\begin{align}
\bar{R}_S(b)&\equiv 2M_B-\frac{4\gamma^2\tilde\Delta M_B^2}{b^3}\,,\quad\bar\Delta\equiv b^2-\bar{R}_S(b)b+a^2\,,\nonumber\\ \bar\Sigma&\equiv\left(b^2+a^2\right)^2-\bar\Delta a^2\sin^2\theta\,.
\end{align}

\section{Absence of closed timelike curves}\label{sec.ccc}
One important feature of the rLQGO is that the closed timelike curves in the classical Kerr spactime naturally disappear. In the classical Kerr spacetime, the closed timelike curves may appear near the ring singularity \cite{RTbook}. This can be understood by considering the norm of the Killing vector $V^\mu$ along the $\varphi$ direction: 
\begin{equation}
	V_\mu V^\mu=g_{\varphi\varphi}\,.
\end{equation}
To approximate the vicinity of the ring singularity inside a Kerr black hole, we consider $\theta=\pi/2+\delta_b$ and expand the norm $V_\mu V^\mu$ with respect to a small quantity $\delta_b\equiv b/a$. We get
\begin{equation}
	V_\mu V^\mu=\frac{aM_B}{\delta_b}+a^2+\mathcal{O}(\delta_b)\,,
\end{equation} 
which can be negative for a small negative $\delta_b$. This leads to the possibilities of having closed timelike curves near the ring singularity. However, in rLQGO, even if we consider the spacetime regions very close to the transition surface where the curvature is expected to be sizable, by expanding $V_\mu V^\mu$ with respect to $\delta_y\equiv y/a$ and assuming $\theta=\pi/2+\delta_y$, we get
\begin{equation}
	V_\mu V^\mu=2\left[a^2\left(\sqrt{\frac{2}{A_\lambda}}-1\right)+A_\lambda M_B^2\right]+\mathcal{O}(\delta_y^2)\,.
\end{equation}
It can be proven that the leading-order term is always non-negative when $|a|\le M_B$. First, for a given value of $a$, the minimum of $I\equiv a^2(\sqrt{2/A_\lambda}-1)+A_\lambda M_B^2$ appears when $\sqrt{2A_\lambda^3}=a^2/M_B^2$. Replacing $a^2/M_B^2$ in $I$ with $\sqrt{2A_\lambda^3}$, one can find that $I$ has a minimum at $a=0$ on which $A=0$ and $I=0$. Therefore, the leading order  of $V_\mu V^\mu$ is always positive for rLQGO and we conclude that there is no closed timelike curve, as opposed to the classical Kerr black hole.

\section{Geodesic equations}\label{sec.geo}
For the classical Kerr black hole metric, there is a hidden symmetry characterized by the existence of the Killing tensor. It can be shown that the geodesic equations allow complete separations of variables using the Carter constant associated with the Killing tensor \cite{Carter:1968ks}. After separating the variables, the geodesic equations can be rewritten in their first-order form. For a rotating spacetime generated from NJA (with an arbitrary $\Psi$), only the photon geodesic equations respect such a symmetry and are separable \cite{Shaikh:2019fpu,Junior:2020lya}. In order for the timelike geodesics to be separable as well, the spacetime metric should fulfill one more condition that the metric function $\Psi$ has to be additively separable \cite{Chen:2019jbs}. The metric function $\Psi$ given in Eq.~\eqref{psichoise} in the rLQGO model is indeed additively separable. Therefore, the geodesic equations (both lightlike and timelike geodesics) allow separations of variables. We will show this in more detail here.

The geodesic equations are described by the following Lagrangian:
\begin{equation}
	\mathcal{L}=\frac{1}{2}g_{\mu\nu}\dot{x}^\mu \dot{x}^\nu\,,\label{geodela}
\end{equation}
where the dot denotes the derivative with respect to $\lambda$, which stands for an affine parameter for a massless particle, or the proper time for a massive object. Because the rLQGO model is stationary and axisymmetric, the geodesic equations contain two constants of motion: the conserved energy $E$ and the conserved azimuthal angular momentum $L_z$. In terms of these constants of motion, one gets the following two geodesic equations in the $t$ and $\varphi$ sectors
\begin{equation}
	\dot{t}=\frac{Eg_{\varphi\varphi}+L_zg_{t\varphi}}{g_{t\varphi}^2-g_{tt}g_{\varphi\varphi}}\,,\qquad
	\dot{\varphi}=-\frac{Eg_{t\varphi}+L_zg_{tt}}{g_{t\varphi}^2-g_{tt}g_{\varphi\varphi}}\,,
\end{equation}
which can be written explicitly as
\begin{align}
	\rho^2\dot{t}=&\frac{\left(a^2+b^2\right)\left[\left(a^2+b^2\right)E-aL_z\right]}{\Delta}+aL_z-a^2E\sin^2\theta\,,\nonumber\\
	\rho^2\dot{\varphi}=&\frac{aE\left(a^2+b^2\right)-a^2L_z}{\Delta}-aE+L_z\csc^2\theta\,.\label{geotphi}
\end{align}

The $y$ and $\theta$ sectors of the geodesic equations can be obtained using the Hamilton-Jacobi approach. The Hamilton-Jacobi equation associated with the Lagrangian \eqref{geodela} reads
\begin{equation}
	\frac{\partial\mathcal{S}}{\partial\lambda}+\mathcal{H}=0\,,\label{HJeq}
\end{equation}
where $\mathcal{S}$ is the Jacobi action. The Hamiltonian $\mathcal{H}$ can be written as $\mathcal{H}=p_\mu p^\mu/2$, where $p_\mu$ is the conjugate momentum:
\begin{equation}
	p_\mu\equiv\frac{\partial\mathcal{L}}{\partial\dot{x}^\mu}=g_{\mu\nu}\dot{x}^\nu=\frac{\partial\mathcal{S}}{\partial x^\mu}\,.\label{conjugatemom}
\end{equation}
Using the above Hamiltonian, the Hamilton-Jacobi equation \eqref{HJeq} can be written explicitly as
\begin{equation}
	\frac{\partial\mathcal{S}}{\partial\lambda}=-\frac{1}{2}g^{\mu\nu}\frac{\partial\mathcal{S}}{\partial x^\mu}\frac{\partial\mathcal{S}}{\partial x^\nu}\,.\label{HJeq2}
\end{equation}
We write the Jacobi action as the following additive form:
\begin{equation}
	\mathcal{S}=\frac{1}{2}\epsilon\lambda-Et+L_z\varphi+S_y(y)+S_\theta(\theta)\,,\label{HJansatz}
\end{equation}
where $\epsilon=0$ ($\epsilon=1$) for lightlike (timelike) geodesics.

Inserting the ansatz \eqref{HJansatz} into the Hamilton-Jacobi equation \eqref{HJeq2} and introducing a decoupling constant (the Carter constant $\mathcal{K}$), one can obtain
\begin{equation}
	\Delta\left(\frac{dS_y}{dy}\right)^2=\frac{\mathcal Y(y)}{\Delta}\,,\quad
	\left(\frac{dS_\theta}{d\theta}\right)^2=\Theta(\theta)\,,
\end{equation}
where
\begin{align}
	\mathcal{Y}(y)&\equiv\left[\left(b^2+a^2\right)E-aL_z\right]^2\nonumber\\&-\Delta\left[\epsilon b^2+\mathcal{K}+\left(L_z-aE\right)^2\right]\,,\\
	\Theta(\theta)&\equiv\mathcal{K}+\cos^2\theta\left(a^2E^2-L_z^2\csc^2\theta-\epsilon a^2\right)\,.
\end{align}
Finally, using the last equality in Eq.~\eqref{conjugatemom}, one can express the $y$ and $\theta$ sectors of the geodesic equations in their first-order form
\begin{align}
	\rho^2\dot{y}&=\pm\sqrt{\mathcal{Y}(y)}\,,\label{geoy}\\
	\rho^2\dot{\theta}&=\pm\sqrt{\Theta(\theta)}\,.\label{geotheta}
\end{align}
Eqs.~\eqref{geotphi}, \eqref{geoy}, and \eqref{geotheta} are the geodesic equations in the rLQGO spacetime and they are indeed separable.\\

\section{Klein-Gordon equation}\label{sec.KG}
Here, we shall show that the KG equation in the rLQGO spacetime allows a separation of variables. The criterion for the separability has been examined in Ref.~\cite{Chen:2019jbs} and it turns out that the rLQGO model satisfies the criterion ($F=G$ specifically). Consider the KG equation of a massive scalar field with a mass $\mu$:
\begin{equation}
	\Box\Phi-\mu^2\Phi=0\,.\label{KGeq}
\end{equation}
Using the rLQGO metric \eqref{finalrotapp}, the KG equation \eqref{KGeq} can be expressed as
\begin{align}
	0=&\,\partial_y\left(\Delta\partial_y\Phi\right)-\frac{a^2}{\Delta}\partial_\varphi^2\Phi+2a\left(1-\frac{a^2+b^2}{\Delta}\right)\partial_t\partial_\varphi\Phi\nonumber\\
	&-\left(b^2+a^2\cos^2\theta\right)\mu^2\Phi-\frac{\left(a^2+b^2\right)^2}{\Delta}\partial_t^2\Phi\nonumber\\&+\frac{1}{\sin\theta}\partial_\theta\left(\sin\theta\partial_\theta\Phi\right)+\frac{1}{\sin^2\theta}\partial_\varphi^2\Phi+a^2\sin^2\theta\partial_t^2\Phi\,.\label{KGequationintermin1}
\end{align}
To proceed, we consider the following field decomposition:
\begin{equation}
	\Phi(t,y,\theta,\varphi)\equiv e^{-i\omega t+im\varphi}\tilde\Phi(y,\theta)\,.
\end{equation}
Using this ansatz, Eq.~\eqref{KGequationintermin1} can be rewritten as
\begin{align}
	&\partial_y\left(\Delta\partial_y\tilde\Phi\right)+\partial_z\left[\left(1-z^2\right)\partial_z\tilde\Phi\right]\nonumber\\
	&+\Bigg[\frac{\left(a^2+b^2\right)^2\omega^2}{\Delta}-\mu^2b^2-\mu^2a^2z^2-\frac{m^2}{1-z^2}\nonumber\\&-a^2\left(1-z^2\right)\omega^2+\frac{a^2m^2}{\Delta}\nonumber\\&+2a\omega m\left(1-\frac{a^2+b^2}{\Delta}\right)\Bigg]\tilde\Phi=0\,,\label{KGequationintermin2}
\end{align}
where we have defined $z\equiv\cos\theta$. Finally, by setting $\tilde\Phi(y,z)=Y(y)Z(z)$ and introducing a separation constant $C$, Eq.~\eqref{KGequationintermin2} can be separated as follows
	\begin{align}
		&\left\{\partial_z\left[\left(1-z^2\right)\partial_z\right]+\left(\omega^2-\mu^2\right)a^2z^2-m^2z^2\left(1-z^2\right)^{-1}-\left(m-a\omega\right)^2+C\right\}Z(z)=0\,,\\
		&\left\{\partial_y\left(\Delta\partial_y\right)+\left[\omega\left(a^2+b^2\right)-am\right]^2\Delta^{-1}-\mu^2b^2-C\right\}Y(y)=0\,.
	\end{align}

\end{document}